\documentclass[conference]{IEEEtran}
\IEEEoverridecommandlockouts

\usepackage{cite}
\usepackage{amsmath,amssymb}
\usepackage{graphicx}
\usepackage{url}
\usepackage{booktabs}
\usepackage{listings}
\usepackage{xcolor}

\begin{document}

\title{Weave: Verified Netlist-to-Schematic Conversion\\ via Layered Graph Layout}

\author{\IEEEauthorblockN{Senol Gulgonul}
\IEEEauthorblockA{Department of Electrical and Electronics Engineering\\
Ostim Technical University, Ankara, Turkey}}

\maketitle

\begin{abstract}
Converting a SPICE netlist into a human-readable schematic is a longstanding problem in electronic design automation: simulators and machine-learning pipelines readily produce netlists, but designers reason about circuits through diagrams. Recent learning-based approaches translate netlists into schematics probabilistically, yet they provide no guarantee that the generated drawing preserves the original connectivity, and their accuracy degrades sharply as circuits grow. We present Weave, a deterministic converter that turns a SPICE netlist into an LTspice \texttt{.asc} schematic using a layered (Sugiyama-style) graph layout, and that certifies every output by a round-trip connectivity check: the generated schematic is re-parsed into a netlist and compared, net for net, against the input. A result is reported as correct only when the two partitions are identical, giving a binary correctness certificate rather than a similarity score. Weave runs entirely client-side as a single dependency-free file and embeds a pin table for 5093 LTspice symbols. On the identical public Circuits-LTSpice test set used by the state-of-the-art LLM converter Schemato (117 circuits, netlisted with LTspice itself), Weave achieves 100\% compilation and 100\% round-trip-verified connectivity equivalence, compared with Schemato's reported 76\% compilation and a graph-edit-distance similarity of 0.35; notably, 73\% of that set exceeds the five-component threshold beyond which Schemato reports losing connectivity accuracy. On a larger and harder corpus, the 3460 netlistable circuits of the official Analog Devices LTspice demo collection, Weave verifies exact connectivity for 88.4\% of circuits, with the remaining failures concentrated in a single, well-characterized class of dense multi-pin power modules.
\end{abstract}


\section{Introduction}

The netlist is the lingua franca of circuit simulation, but it is not how engineers understand circuits. Designers rely on schematics to recognize structure, trace signal flow, and debug behavior~\cite{yang2025review}. As machine-learning methods increasingly generate circuits directly as netlists, the gap between machine-produced netlists and human-interpretable schematics has become a practical bottleneck: to bring domain experts into the loop, netlists must be turned back into readable diagrams quickly and accurately~\cite{schemato2025}.

Automatic schematic generation (ASG) is a mature field. The recent review by Yang et al.~\cite{yang2025review} organizes ASG layout into two families: heuristic-based methods and, more recently, deep-reinforcement-learning-based methods. The heuristic line is classical and well understood. Swinkels and Hafer~\cite{swinkels1990} framed schematic layout as an expert-system problem with horizontal and vertical ordering; Jehng et al.~\cite{jehng1991asg} introduced the ASG value-propagation layout; Arsintescu~\cite{arsintescu1996} placed symmetric pairs on a grid and minimized wire length and bends; Wu~\cite{wu2009} and others advanced function-block placement along the signal-flow direction. Naveen and Raghunathan~\cite{naveen1993n2s} built an early netlist-to-schematic generator with channel routing, and Frezza and Levitan~\cite{frezza1993spar} combined placement and routing in a single system. These methods are interpretable and fast, but, as the review notes, they typically settle for a good rather than an optimal arrangement~\cite{yang2025review}.

The modern wave is learning-based. Hsu and Lin~\cite{hsu2022} generate analog schematics through building-block classification and reinforcement learning, optimizing an aesthetic reward. Most directly related to our work, Schemato~\cite{schemato2025} fine-tunes a large language model to translate netlists into LTspice \texttt{.asc} files, reporting up to 76\% compilation success and outperforming general-purpose LLMs. These approaches produce visually plausible schematics, but they share two structural limitations that the review and the Schemato paper both make explicit. First, they degrade with scale: the review observes that the reinforcement-learning approach of Hsu and Lin does not exceed roughly 40 components~\cite{yang2025review}, and Schemato itself reports that ``for circuits with more than 5 components, Schemato struggles to generate schematics with accurate connectivity''~\cite{schemato2025}. Second, and more fundamentally, they offer no guarantee of correctness: their quality is measured by continuous similarity scores such as graph edit distance (GED)~\cite{abuaisheh2015ged} and structural similarity (MSSIM)~\cite{wang2004ssim}, and a substantial fraction of their outputs do not even compile.

We take a different position. We argue that for a converter to be useful in an engineering workflow, connectivity correctness should be \emph{guaranteed by construction and verified}, not approximated and scored. We present \textbf{Weave}, a deterministic netlist-to-schematic converter with three contributions:

\begin{enumerate}
\item \textbf{A deterministic, verification-first pipeline.} Weave revives the classical layered-flow layout line and rebuilds it on a modern industrial graph engine (elkjs, an implementation of the Sugiyama framework), combined with a small set of placement patterns for feedback loops, divider legs, and shunts. No training, no GPU, no probabilistic sampling.
\item \textbf{Round-trip connectivity verification.} Every generated \texttt{.asc} is parsed back into a netlist and compared, net for net, against the input. A result is accepted only when the two connectivity partitions are identical---a binary certificate equivalent to $\mathrm{GED}=1.0$, produced at generation time. To our knowledge, no prior netlist-to-schematic method certifies connectivity in this way.
\item \textbf{A safe-mode ladder.} When the default layout leaves a connectivity gap, Weave retries with progressively simpler layout modes down to pure graph layout, keeping the first result the verifier accepts. Because the verifier gates every rung, a reported match is never wrong---only plainer.
\end{enumerate}

Weave runs entirely in the browser as a single dependency-free file and ships with a command-line tool for batch conversion and reproducible benchmarking.\footnote{The tool is available at \url{https://senolgulgonul.github.io/weave}; source code, the CLI, and benchmark-reproduction scripts at \url{https://github.com/senolgulgonul/weave}.} On the identical Circuits-LTSpice test set used by Schemato, Weave achieves 100\% compilation and 100\% round-trip-verified connectivity, against Schemato's 76\% compilation and 0.35 GED similarity, on a set where 73\% of circuits lie beyond Schemato's stated five-component limit.

\section{Related Work}

\subsection{Heuristic layout for schematics}
The heuristic ASG line established the layered-flow paradigm that Weave builds on. The layout is typically split into logical (global) and geometric (local) phases~\cite{yang2025review}. In the logical phase, a horizontal ordering assigns components to columns following left-to-right signal flow, detecting and removing feedback loops so the remainder is acyclic; a vertical ordering then assigns rows, classically by a barycenter or value-propagation rule~\cite{swinkels1990,jehng1991asg}. Arsintescu~\cite{arsintescu1996} added grid snapping and mirror placement of symmetric pairs with wire-length and bend minimization, while later work placed function blocks incrementally along the flow direction~\cite{wu2009}. Simulated annealing was applied to the same problem as a global optimizer~\cite{lee1989sa}, but, as the review notes, its slow convergence on large schematics limited its adoption~\cite{yang2025review}.

Routing in this line follows its own taxonomy~\cite{yang2025review}: sequential methods route nets one at a time, from maze search~\cite{abel1972} through A*-guided search as used in SPAR~\cite{hart1968astar,frezza1993spar} to constant-time pattern routing~\cite{kastner2000pattern}, with rectilinear Steiner trees for multi-terminal nets~\cite{lee1992aesthetic}; simultaneous methods trade running time for global optimality. Weave does not implement a separate router: orthogonal edge routing is performed jointly with placement by the layered engine, and the wires emitted by placement patterns are direct, net-tagged connections that the verifier audits. These classical ideas map almost directly onto Weave's pipeline---layered assignment, feedback handling as a placement pattern, grid snapping, and orthogonal routing---but Weave replaces the hand-built layout heuristics with a general Sugiyama engine and, crucially, adds a correctness certificate the classical line never had.

\subsection{Learning-based schematic generation}
Hsu and Lin~\cite{hsu2022} treat placement as a reinforcement-learning problem, rewarding aesthetic quality (crossings, bends, wire length). For PCB reverse engineering, the AEM-PCB reverser~\cite{yang2024aem} drives a reinforcement-learning agent with an aesthetic evaluation metric designed to align with designers' subjective judgments, addressing the historical lack of objective quality standards for PCB schematics~\cite{yang2025review}. Schemato~\cite{schemato2025} fine-tunes an LLM on netlist-schematic pairs to emit \texttt{.asc} directly. All three report strong results within their regimes, but all are bounded by scale and by the absence of a correctness guarantee. The review~\cite{yang2025review} attributes the scale limit to the exponential growth of the search or generation space with circuit density, and suggests hierarchical or heuristic pruning as the way forward---precisely the role played by Weave's placement patterns and safe-mode ladder. Schemato additionally reports a data-scarcity limitation: more than half of its training samples contain components seen fewer than ten times, so it cannot reliably place unfamiliar parts~\cite{schemato2025}. Weave sidesteps this entirely with a deterministic symbol table and a generic-block fallback, so connectivity is preserved even for parts whose exact symbol is unknown.

\subsection{Evaluation}
Learning-based converters are evaluated with similarity and compilation metrics: GED~\cite{abuaisheh2015ged} for topological similarity, MSSIM~\cite{wang2004ssim} for visual similarity, and a compilation success rate (CSR) for whether the output opens in the target tool~\cite{schemato2025}. On the aesthetics side, the AEM~\cite{yang2024aem} quantifies layout and routing quality against designer judgment. These are appropriate for probabilistic methods, but they leave correctness as a matter of degree. Weave instead reports a binary, exact connectivity check; a Weave ``match'' corresponds to the best case those metrics can represent ($\mathrm{GED}=1.0$), obtained as a guarantee rather than an average. We view the two axes as complementary: aesthetic metrics such as AEM measure how pleasant a schematic is, while round-trip verification certifies that it is \emph{right}.

\section{Method}

\subsection{Overview}
\begin{figure}[t]
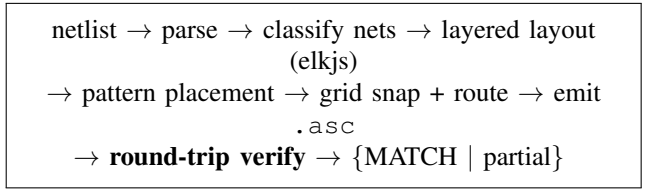

\centering
\framebox[0.95\columnwidth]{\parbox{0.9\columnwidth}{\centering\vspace{1mm}
netlist $\rightarrow$ parse $\rightarrow$ classify nets $\rightarrow$ layered layout (elkjs)\\ $\rightarrow$ pattern placement $\rightarrow$ grid snap + route $\rightarrow$ emit \texttt{.asc}\\ $\rightarrow$ \textbf{round-trip verify} $\rightarrow$ \{MATCH $\mid$ partial\}
\vspace{1mm}}}
\caption{The Weave pipeline. The final round-trip verification certifies that the emitted schematic preserves the input connectivity.}
\label{fig:pipeline}
\end{figure}

Weave converts a SPICE netlist to an LTspice \texttt{.asc} schematic in a short, deterministic pipeline: parse, classify nets, place, route, emit, and verify (Fig.~\ref{fig:pipeline}). Each stage has a single responsibility, and the final stage certifies the result.

\subsection{Parsing and net classification}
The parser covers the standard SPICE element set: passives and sources (R, C, L, V, I, D), active devices (Q, M with substrate collapse, J), subcircuits (X with parameter tails), dependent sources (E, G, F, H), behavioral sources (B), switches (S, W), transmission lines (T), coupling (K), and special-function/digital devices (A), with SPICE line continuations joined. Each net is then classified as ground, supply rail, or signal. Ground and rail nets are realized as flags (local connections) rather than drawn wires, which keeps the signal structure legible; only signal nets enter the layout graph.

\subsection{Layered layout}
Signal components become nodes in a graph laid out by elkjs using its layered (Sugiyama) algorithm with left-to-right flow and orthogonal routing. Node sizes encode the symbol body plus reserved space for pin escapes and flag stubs, all snapped to LTspice's 16-unit grid so that ports land on grid and pin coordinates are exact. Elkjs guarantees non-overlapping node boxes and orthogonal edge routes; Weave translates the resulting coordinates and edge sections into \texttt{.asc} symbols and wires.

\subsection{Placement patterns}
Certain structures are handled outside the layout graph, as placement patterns, because forcing them into the graph degrades the main signal chain: local and far feedback around amplifiers, divider legs to ground or rail, hanging shunts, and supply corners. Each pattern places its element relative to already-placed nodes and emits the connecting wires directly. This mirrors the classical heuristic line~\cite{swinkels1990,arsintescu1996}, but every pattern's wires are tagged with their net so the verifier can audit them.

\subsection{Symbol table and generic fallback}
Weave embeds a pin table for 5093 LTspice symbols, storing only what layout needs: pin offsets and SpiceOrder indices. When a netlist calls a known part whose symbol pin count differs from the call, or a part absent from the table, Weave substitutes a synthetic rectangular block sized to the netlist's pin count. The glyph is generic, but the name and connectivity are exact, so the schematic still opens in LTspice and still passes verification. This is the deterministic counterpart to the data-scarcity problem that limits learned models~\cite{schemato2025}.

\subsection{Round-trip verification}
The verifier is the core of Weave's correctness claim. Given the emitted \texttt{.asc}, it reconstructs a connectivity graph by unioning wire endpoints, flags, and symbol pins that share coordinates into nets, then partitions both the generated schematic and the original netlist into nets and compares the partitions. If they are identical, connectivity is preserved and the result is a \emph{match}; otherwise the verifier reports exactly which nets differ. This check is binary and exact. It is equivalent to a graph-edit-distance score of $1.0$ between the schematic's connectivity graph and the netlist, but it is produced as a certificate at generation time rather than estimated as a similarity.

\subsection{Safe-mode ladder}
A single circuit may not lay out cleanly with all patterns active. Rather than fail, Weave climbs a ladder of increasingly conservative modes, disabling placement patterns one at a time down to pure elkjs layout and, at the extreme, widening channel spacing. It applies the verifier at each rung and keeps the first result that passes. Because correctness is gated by the verifier, a match from any rung is fully correct; higher rungs simply trade some visual compactness for a clean, verifiable result. If any rung produces a connectivity-preserving schematic, Weave returns it, and the outcome is always labeled honestly.

\section{Experimental Setup}

\subsection{Benchmark against Schemato}
To compare directly with the state of the art, we evaluate on the exact public test set used by Schemato: the Circuits-LTSpice repository\footnote{\url{https://github.com/mick001/Circuits-LTSpice}}. We follow Schemato's methodology precisely. Starting from the repository's \texttt{.asc} files, we generate netlists with LTspice itself (\texttt{XVIIx64.exe -netlist}), not with Weave's own extractor, so there is no circularity: the input to Weave comes from LTspice, exactly as Schemato's input did. Circuits whose symbols are missing from the installed library cannot be netlisted and are omitted, as in Schemato. With LTspice XVII 17.0.36.0 this yields 117 netlists, identical to the count Schemato reports. Because both methods are evaluated on the same 117 circuits derived by the same tool, the denominator is identical.

For each netlist, Weave produces an \texttt{.asc} (climbing the safe-mode ladder) and the round-trip verifier reports MATCH or partial. We report compilation rate (a valid \texttt{.asc} produced) and the round-trip-verified MATCH rate (exact connectivity equivalence). Weave can be run either as a browser application at \url{https://senolgulgonul.github.io/weave/}, where a netlist is pasted and the verified \texttt{.asc} downloaded interactively, or through the command-line tool used for the batch results reported here.

\subsection{Large-corpus evaluation}
To probe scale and to characterize failure honestly, we additionally evaluate on the official LTspice demo-circuit collection published by Analog Devices~\cite{adidemo}: 3610 \texttt{.asc} circuits retrieved on July 1, 2026. Applying the same protocol as Sec.~IV-A (netlisting with LTspice XVII 17.0.36.0; circuits with symbols missing from the installed library are omitted) yields a corpus of 3460 netlists. Table~\ref{tab:corpus} summarizes the corpus. It is substantially larger than any test set in the related work, its circuits are markedly denser (mean 14.4 elements, median 12, up to 138), and it contains many dense multi-pin power modules that are absent from the generic-component Circuits-LTSpice set. Because the collection is public and the protocol is fully specified, other converters can be evaluated on the identical corpus for direct comparison.

\begin{table}[t]
\caption{The large evaluation corpus: LTspice demo circuits~\cite{adidemo}, netlisted with LTspice XVII 17.0.36.0.}
\label{tab:corpus}
\centering
\begin{tabular}{lr}
\toprule
Source \texttt{.asc} circuits (retrieved 2026-07-01) & 3610 \\
Netlistable with XVII 17.0.36.0 (evaluated corpus) & 3460 \\
Total circuit elements & 49\,791 \\
Elements per circuit (mean / median / max) & 14.4 / 12 / 138 \\
\midrule
Circuits with 2--5 elements & 261 \\
Circuits with 6--10 elements & 1274 \\
Circuits with 11--20 elements & 1274 \\
Circuits with 21--50 elements & 615 \\
Circuits with 51+ elements & 36 \\
\midrule
Round-trip verified MATCH & 3060 (88.4\%) \\
Partial (connectivity gaps, downloadable) & 285 (8.2\%) \\
Conversion error & 115 (3.3\%) \\
\bottomrule
\end{tabular}
\end{table}

\section{Results}

\subsection{Head-to-head on Circuits-LTSpice}
Table~\ref{tab:schemato} reports the comparison. On the identical 117-circuit set, Weave produces a valid schematic for every circuit and every one is a round-trip-verified MATCH: 100\% compilation and 100\% connectivity equivalence. Schemato's best reported figures on the same set are 76\% compilation and a CSR-scaled GED similarity of 0.35~\cite{schemato2025}; GPT-4o reaches 63\% compilation.

\begin{figure}[!htb]
\centering
\begin{minipage}[c]{0.44\columnwidth}
\begin{lstlisting}
T1 N002 0 N003 0
+ Td=50n Z0=50
V1 N001 0 AC 1
RS N002 N001 50
RL N003 0 50
\end{lstlisting}
\centering \footnotesize (a) input netlist
\end{minipage}\hfill
\begin{minipage}[c]{0.52\columnwidth}
\centering
\includegraphics[width=\linewidth]{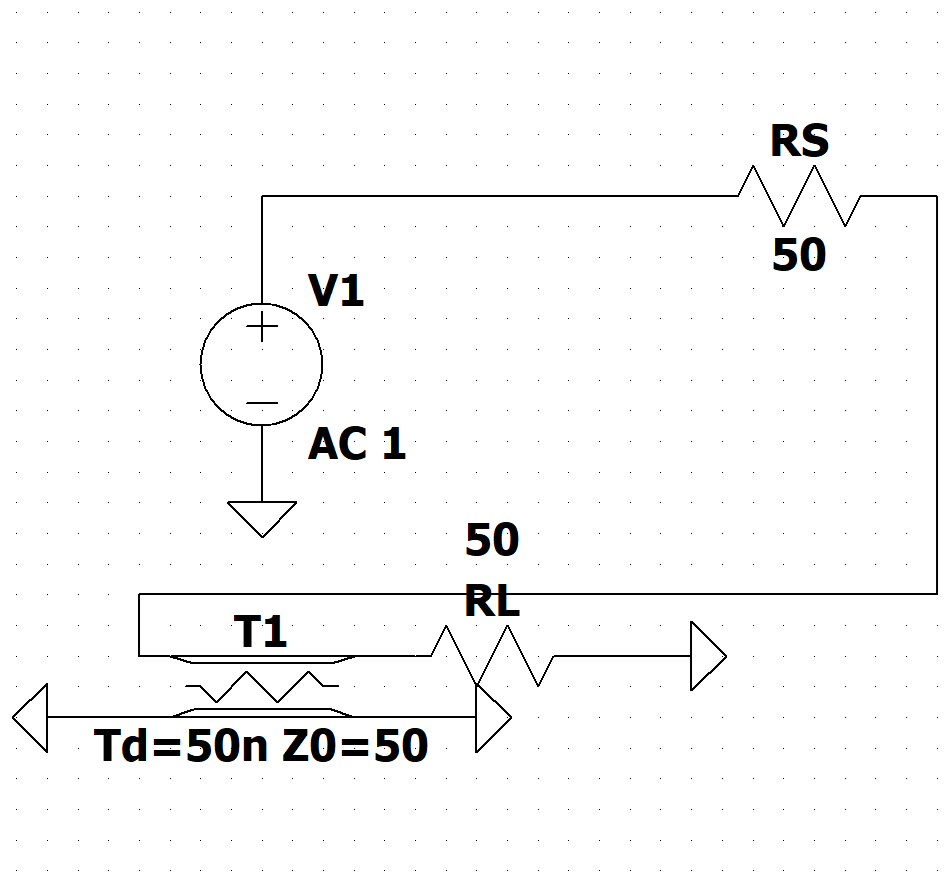}\\ \footnotesize (b) Weave output
\end{minipage}
\caption{Transformer circuit: input netlist (a) and Weave output (b).}
\label{fig:transformer}
\end{figure}

\begin{figure}[!htb]
\centering
\begin{minipage}[c]{0.44\columnwidth}
\begin{lstlisting}
V1 N001 0 AC 1
C1 N002 N001 100n
R1 N002 Vout 10k
R2 Vout 0 10k
C2 Vout 0 100n
\end{lstlisting}
\centering \footnotesize (a) input netlist
\end{minipage}\hfill
\begin{minipage}[c]{0.52\columnwidth}
\centering
\includegraphics[width=\linewidth]{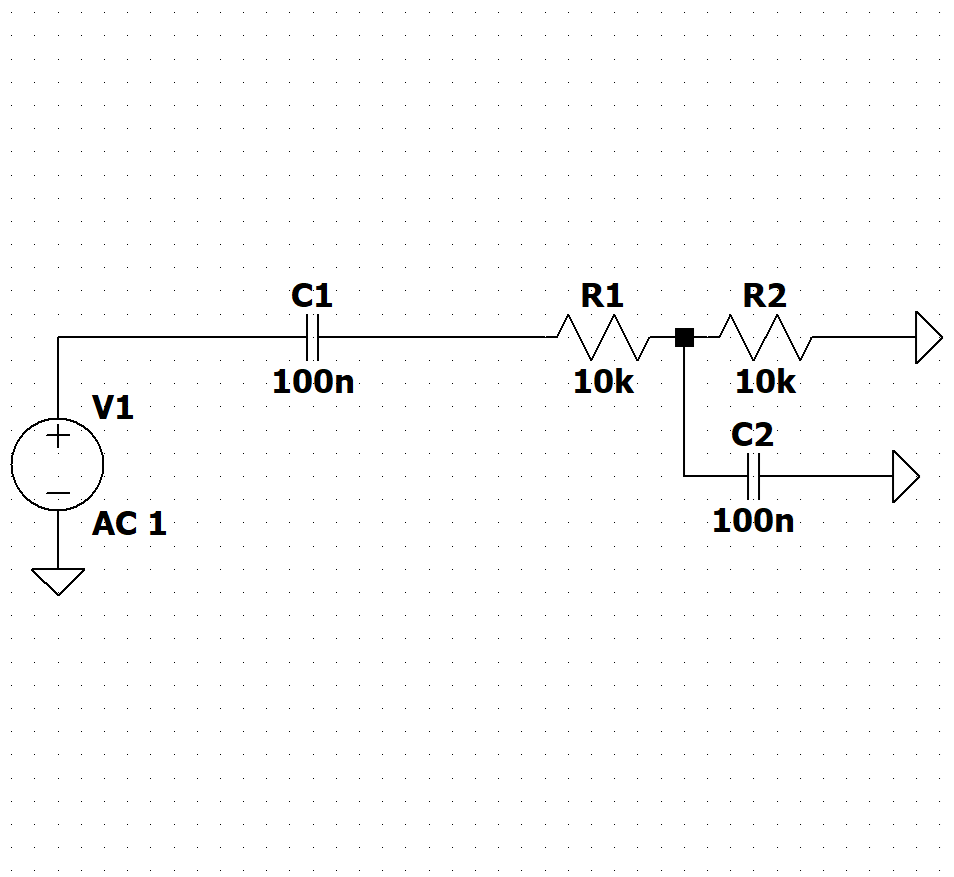}\\ \footnotesize (b) Weave output
\end{minipage}
\caption{RC band-pass filter: input netlist (a) and Weave output (b).}
\label{fig:rcband}
\end{figure}

Figures~\ref{fig:transformer} and~\ref{fig:rcband} show two circuits from this set used as qualitative examples in~\cite{schemato2025}: a transformer circuit and an RC band-pass filter, each with its input netlist and the schematic Weave generates from it. The transformer netlist in Fig.~\ref{fig:transformer} also illustrates the SPICE \texttt{+} line continuation, which the parser joins automatically. Both outputs are round-trip-verified (exact connectivity). Schemato reports that GPT-4o fails to produce a compilable schematic for the transformer circuit across all prompts~\cite{schemato2025}, whereas Weave produces a verified schematic with clean left-to-right signal flow for both. Each conversion can be reproduced either with the command-line tool or interactively in the browser at \url{https://senolgulgonul.github.io/weave/} by pasting the netlist.

\begin{table}[!b]
\caption{Comparison on the identical Circuits-LTSpice test set (117 circuits, netlisted by LTspice). Schemato and GPT-4o figures are as reported in~\cite{schemato2025}.}
\label{tab:schemato}
\centering
\begin{tabular}{lccc}
\toprule
Metric & Schemato & GPT-4o & \textbf{Weave} \\
\midrule
Compilation rate & 76\% & 63\% & \textbf{100\%} \\
Connectivity & GED 0.35 & GED 0.23 & \textbf{100\% exact} \\
Guarantee & none & none & \textbf{per-output} \\
Infrastructure & 8$\times$GPU & API & \textbf{browser} \\
\bottomrule
\end{tabular}
\end{table}

\begin{figure}[!htb]
\centering
\includegraphics[width=0.80\columnwidth]{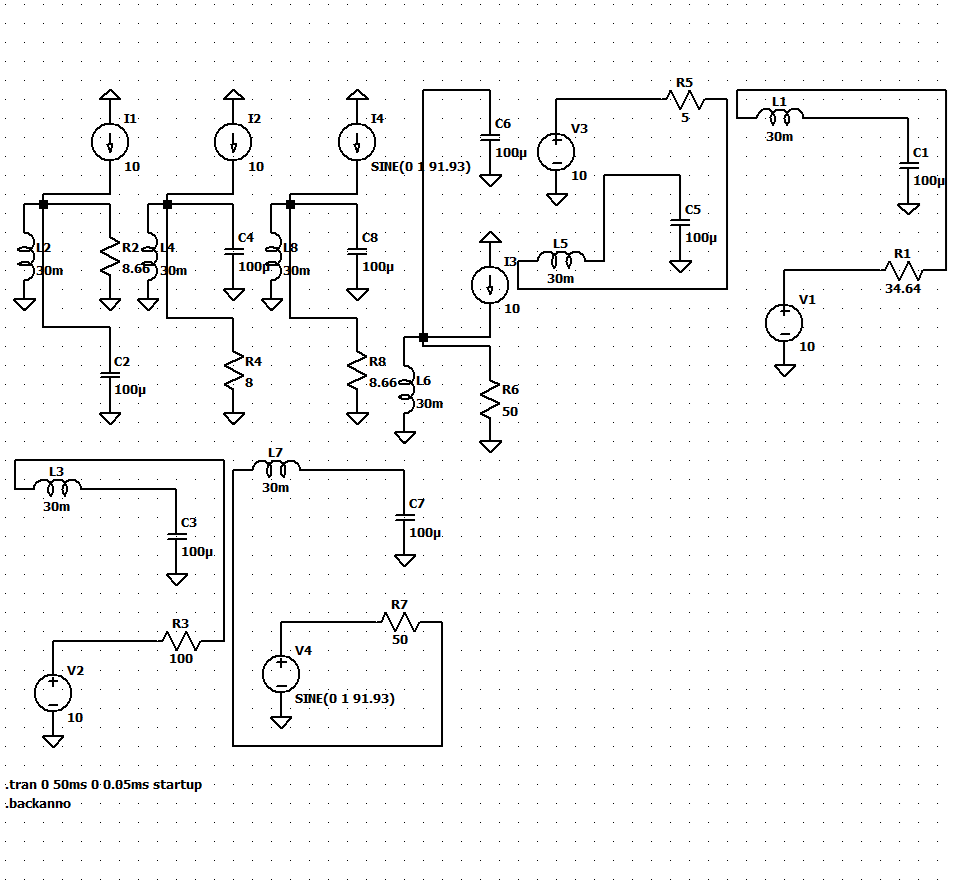}
\caption{Weave output for a 32-component RLC step-response network.}
\label{fig:hard}
\end{figure}

The result is not an artifact of an easy set. The 117 circuits average 11 components (median 9, maximum 39), and 86 of them (73\%) exceed five components---exactly the region where Schemato reports losing connectivity accuracy~\cite{schemato2025}. Weave returns a verified match for all of them, including the largest circuits (37--39 components: a three-phase supply and a common-emitter amplifier design). Figures~\ref{fig:hardnet} and~\ref{fig:hard} show one such circuit, a 32-component RLC step-response network, more than six times the five-component threshold: Fig.~\ref{fig:hardnet} lists the input netlist and Fig.~\ref{fig:hard} the schematic Weave generates from it; the layout remains balanced and the connectivity is round-trip-verified (exact).

\begin{figure}[!htb]
\centering
\begin{minipage}[t]{0.48\columnwidth}
\begin{lstlisting}
V1 N004 0 10
L1 N005 N006 30m
R1 N004 N005 34.64
C1 N006 0 100u
L2 N007 0 30m
R2 N007 0 8.66
C2 N007 0 100u
I1 0 N007 10
V2 N009 0 10
L3 N010 N011 30m
R3 N009 N010 100
C3 N011 0 100u
L4 N012 0 30m
R4 N012 0 8
C4 N012 0 100u
I2 0 N012 10
V3 N013 0 10
\end{lstlisting}
\end{minipage}\hfill
\begin{minipage}[t]{0.48\columnwidth}
\begin{lstlisting}
L5 N014 N015 30m
R5 N013 N014 5
C5 N015 0 100u
L6 N016 0 30m
R6 N016 0 50
C6 N016 0 100u
I3 0 N016 10
V4 N001 0 SINE(0 1 91.93)
L7 N002 N003 30m
R7 N001 N002 50
C7 N003 0 100u
L8 N008 0 30m
R8 N008 0 8.66
C8 N008 0 100u
I4 0 N008 SINE(0 1 91.93)
.tran 0 50ms 0 0.05ms
+ startup
\end{lstlisting}
\end{minipage}
\caption{Input netlist of the 32-component RLC step-response network of Fig.~\ref{fig:hard} (annotation comments omitted; the micro suffix is written in its ASCII form \texttt{u}).}
\label{fig:hardnet}
\end{figure}

We confirmed that the verifier is not vacuous: deleting a wire from a generated schematic causes the verifier to report a mismatch (nets break apart), and altering a net in the input netlist likewise produces a mismatch. The 100\% therefore reflects genuine connectivity equivalence, not a verifier that accepts everything.

\subsection{Large-corpus behavior}
On the 3460-circuit corpus, Weave verifies exact connectivity for 3060 circuits (88.4\%); 285 (8.2\%) are partial results with reported connectivity gaps, and 115 (3.3\%) fail to convert (Table~\ref{tab:corpus}). The failures are highly concentrated: the great majority of partials involve dense multi-pin power modules (large regulators with many repeated nets), which form a single, well-characterized class rather than a scattered set of errors, and most conversion errors stem from parts newer than the installed LTspice version, whose symbols are absent from the table. This corpus is deliberately harder than Circuits-LTSpice---it is dominated by exactly the components that are rare in the generic-component benchmark---and it delineates precisely where the current layout reaches its limit.

\subsection{Discussion}
The contrast with learning-based converters is structural rather than incremental. Schemato's difficulty beyond five components and its data-scarcity limitation are inherent to a probabilistic model that must learn both connectivity and symbol geometry from examples~\cite{schemato2025}. Weave separates these concerns: connectivity is constructed deterministically from the netlist and then verified, while symbol geometry comes from a fixed table with a generic fallback. The cost of this design is aesthetic: Weave optimizes for verified connectivity first, and does not yet report the crossing, bend, or visual-similarity metrics that the learning-based line emphasizes~\cite{yang2025review,hsu2022}. We regard this as an honest scoping of the contribution: a converter that is provably correct on connectivity is a stronger primitive to build aesthetics on than an aesthetically tuned one whose connectivity is uncertain.

\section{Conclusion}

We presented Weave, a deterministic netlist-to-schematic converter that certifies every output by round-trip connectivity verification and, through a safe-mode ladder, guarantees an always-correct, as-clean-as-possible result. On the identical Circuits-LTSpice benchmark used by the state-of-the-art LLM converter, Weave achieves 100\% compilation and 100\% verified connectivity against 76\% and 0.35 GED, on a set where most circuits exceed the scale at which the learned model degrades.

Two limitations point to the same next step. Connectivity failures concentrate in a single class of dense multi-pin power modules, and, separately, deep cascade topologies (such as multi-stage amplifiers) are stretched by the layered layout into long chains that are correct but not compact. Both are addressed by a box-contract layout in which feedback, far-feedback, and block slots enter the graph as fixed-size nodes so that all routing, including cascades and dense modules, is optimized jointly---a direction consistent with the hierarchical, partition-based generation that the review identifies as the way to scale ASG to dense circuits~\cite{yang2025review,hong2023graphclusnet}. Adding the aesthetic metrics emphasized by the learning-based line, and a direct comparison on those metrics, is the other clear direction. We believe the broader lesson generalizes beyond LTspice: for netlist-to-schematic conversion, correctness by construction with verification is both achievable and, on connectivity, decisively stronger than learned approximation.

\bibliographystyle{IEEEtran}
\bibliography{references}

\begin{thebibliography}{10}
\providecommand{\url}[1]{#1}
\csname url@samestyle\endcsname
\providecommand{\newblock}{\relax}
\providecommand{\bibinfo}[2]{#2}
\providecommand{\BIBentrySTDinterwordspacing}{\spaceskip=0pt\relax}
\providecommand{\BIBentryALTinterwordstretchfactor}{4}
\providecommand{\BIBentryALTinterwordspacing}{\spaceskip=\fontdimen2\font plus
\BIBentryALTinterwordstretchfactor\fontdimen3\font minus
  \fontdimen4\font\relax}
\providecommand{\BIBforeignlanguage}[2]{{%
\expandafter\ifx\csname l@#1\endcsname\relax
\typeout{** WARNING: IEEEtran.bst: No hyphenation pattern has been}%
\typeout{** loaded for the language `#1'. Using the pattern for}%
\typeout{** the default language instead.}%
\else
\language=\csname l@#1\endcsname
\fi
#2}}
\providecommand{\BIBdecl}{\relax}
\BIBdecl

\bibitem{yang2025review}
J.~Yang, K.~Qiao, J.~Chen, C.~Chen, L.~Guo, and B.~Yan, ``A review of automatic
  schematic generation techniques and their application to printed circuit
  boards,'' \emph{Frontiers of Information Technology \& Electronic
  Engineering}, vol.~26, no.~9, pp. 1534--1550, 2025.

\bibitem{schemato2025}
R.~Matsuo, S.~Uhlich, A.~Venkitaraman, A.~Bonetti, C.-Y. Hsieh, A.~Momeni,
  L.~Mauch, A.~Capone, E.~Ohbuchi, and L.~Servadei, ``Schemato -- an {LLM} for
  netlist-to-schematic conversion,'' in \emph{Proc. ACM/IEEE Int. Workshop on
  Machine Learning for CAD (MLCAD)}, 2025.

\bibitem{swinkels1990}
G.~M. Swinkels and L.~Hafer, ``Schematic generation with an expert system,''
  \emph{IEEE Trans. Computer-Aided Design of Integrated Circuits and Systems},
  vol.~9, no.~12, pp. 1289--1306, 1990.

\bibitem{jehng1991asg}
Y.-S. Jehng, L.-G. Chen, and T.-M. Parng, ``{ASG}: automatic schematic
  generator,'' \emph{Integration, the VLSI Journal}, vol.~11, no.~1, pp.
  11--27, 1991.

\bibitem{arsintescu1996}
B.~G. Arsintescu, ``A method for analog circuits visualization,'' in
  \emph{Proc. Int. Conf. on Computer Design (ICCD): VLSI in Computers and
  Processors}, 1996, pp. 454--459.

\bibitem{wu2009}
Y.~P. Wu, ``Novel method of analog circuit schematic synthesis,'' in \emph{IEEE
  8th Int. Conf. on ASIC}, 2009, pp. 1209--1212.

\bibitem{naveen1993n2s}
B.~Naveen and K.~Raghunathan, ``An automatic netlist-to-schematic generator,''
  \emph{IEEE Design \& Test of Computers}, vol.~10, no.~1, pp. 36--41, 1993.

\bibitem{frezza1993spar}
S.~T. Frezza and S.~P. Levitan, ``{SPAR}: a schematic place and route system,''
  \emph{IEEE Trans. Computer-Aided Design of Integrated Circuits and Systems},
  vol.~12, no.~7, pp. 956--973, 1993.

\bibitem{hsu2022}
H.-Y. Hsu and M.~P.-H. Lin, ``Automatic analog schematic diagram generation
  based on building block classification and reinforcement learning,'' in
  \emph{Proc. ACM/IEEE Workshop on Machine Learning for CAD (MLCAD)}, 2022, pp.
  43--48.

\bibitem{abuaisheh2015ged}
Z.~Abu-Aisheh, R.~Raveaux, J.-Y. Ramel, and P.~Martineau, ``An exact graph edit
  distance algorithm for solving pattern recognition problems,'' in \emph{Int.
  Conf. on Pattern Recognition Applications and Methods (ICPRAM)}, 2015.

\bibitem{wang2004ssim}
Z.~Wang, A.~C. Bovik, H.~R. Sheikh, and E.~P. Simoncelli, ``Image quality
  assessment: from error visibility to structural similarity,'' \emph{IEEE
  Trans. Image Processing}, vol.~13, no.~4, pp. 600--612, 2004.

\bibitem{lee1989sa}
T.~D. Lee and L.~P. McNamee, ``Structure optimization in logic schematic
  generation,'' in \emph{IEEE Int. Conf. on Computer-Aided Design (ICCAD)
  Digest of Technical Papers}, 1989, pp. 330--333.

\bibitem{abel1972}
L.~C. Abel, ``On the ordering of connections for automatic wire routing,''
  \emph{IEEE Trans. Computers}, vol. C-21, no.~11, pp. 1227--1233, 1972.

\bibitem{hart1968astar}
P.~E. Hart, N.~J. Nilsson, and B.~Raphael, ``A formal basis for the heuristic
  determination of minimum cost paths,'' \emph{IEEE Trans. Systems Science and
  Cybernetics}, vol.~4, no.~2, pp. 100--107, 1968.

\bibitem{kastner2000pattern}
R.~Kastner, E.~Bozorgzadeh, and M.~Sarrafzadeh, ``Predictable routing,'' in
  \emph{IEEE/ACM Int. Conf. on Computer-Aided Design (ICCAD)}, 2000, pp.
  110--113.

\bibitem{lee1992aesthetic}
Lee and McNamee, ``Aesthetic routing for transistor schematics,'' in
  \emph{IEEE/ACM Int. Conf. on Computer-Aided Design (ICCAD)}, 1992, pp.
  35--38.

\bibitem{yang2024aem}
J.~Yang, K.~Qiao, S.~H. Shi \emph{et~al.}, ``{AEM-PCB} reverser: circuit
  schematic generation in {PCB} reverse engineering using reinforcement
  learning based on aesthetic evaluation metric,'' \emph{IEEE Trans.
  Computer-Aided Design of Integrated Circuits and Systems}, vol.~43, no.~5,
  pp. 1608--1612, 2024.

\bibitem{adidemo}
{Analog Devices}, ``{LTspice} demo circuits,''
  \url{https://www.analog.com/en/resources/design-tools-and-calculators/ltspice-simulator/lt-spice-demo-circuits.html},
  accessed: Jul. 1, 2026.

\bibitem{hong2023graphclusnet}
X.~N. Hong, T.~Lin, Y.~Q. Shi \emph{et~al.}, ``{GraphClusNet}: a hierarchical
  graph neural network for recovered circuit netlist partitioning,'' \emph{IEEE
  Trans. Artificial Intelligence}, vol.~4, no.~5, pp. 1199--1213, 2023.

\end{thebibliography}

\end{document}